\documentstyle[12pt]{JHEP3}

\renewcommand\O{{\mathcal{O}}}

\newcommand{\be}[1]{ \begin{equation}\label{#1} }
\newcommand{\ee}{\end{equation}}
\newcommand{\ben}[1]{\begin{eqnarray}\label{#1} }
\newcommand{\een}{\end{eqnarray}}
\newcommand{\eq}[1]{(\ref{#1})}
\def\ZZZ{{\hskip-3pt\hbox{ Z\kern-1.6mm Z}}}
\def\zzz{{\hskip-3pt\hbox{ z\kern-1mm z}}}

\newcommand{\p}{\partial}

\newcommand{\D}{\Delta}

\newcommand{\refb}[1]{(\ref{#1})}

\def\one{{\hbox{ 1\kern-.8mm l}}}
\def\zero{{\hbox{ 0\kern-1.5mm 0}}}

\title{On Representations and Correlation Functions of Galilean Conformal Algebras}

\author{Arjun Bagchi, Ipsita Mandal\\
$\;$ Harish-Chandra Research Institute,\\
$\;$ Chhatnag Road, Jhusi,\\
$\;$ Allahabad 211019, India\\
$\;$\email{arjun, ipsita@hri.res.in}
}

\abstract{Galilean Conformal Algebras (GCA) have been recently proposed as a different non-relativistic limit of the AdS/CFT conjecture. In this note, we look at the representations of the GCA. We also construct explicitly the two and three point correlators in this non-relativistic limit of CFT and comment on the differences with the relativistic case and also the more studied Schrodinger group.}

\preprint{HRI/ST/0910}

\begin{document}

\baselineskip 3.5ex

\section{Introduction}

The non-relativistic limit of the AdS/CFT conjecture \cite{Maldacena:1997re} has recently received a lot of attention. The motivation has mainly been studying real-life systems in condensed matter physics via the gauge-gravity duality\footnote{See \cite{Hartnoll:2009sz} for a recent review of AdS/Condensed Matter Theory (AdS/CMT) correspondence.}. It was pointed out in \cite{Nishida:2007pj} that the Schr\"{o}dinger symmetry group \cite{Hagen:1972pd, Niederer:1972zz}, a non-relativistic version of conformal symmetry, is relevant to the study of cold atoms. A gravity dual possessing these symmetries was then proposed in \cite{Son:2008ye, Balasubramanian:2008dm} (see also \cite{Goldberger:2008vg, Barbon:2008bg} for a somewhat different bulk realization). Embeddings in string theory were first looked at in \cite{Herzog:2008wg, Maldacena:2008wh, Adams:2008wt}. For a more exhaustive list of references on the current work on Schr\"{o}dinger algebra, we refer the reader to \cite{Bagchi:2009my}.
There also might be possibly interesting tractable sectors of the parent conjecture, like the BMN limit \cite{BMN}, which emerge when we look at the non-relativistic limits.

Recently, the study of a different non-relativistic limit was initiated in \cite{Bagchi:2009my}, where the authors proposed to study a non-relativistic conformal symmetry obtained by a parametric contraction of the relativistic conformal group. 
The process of group contraction of the relativistic conformal group $SO(d+1,2)$ in $d+1$ space-time dimensions, leads in $d=3$ to a fifteen parameter group (like the parent $SO(4,2)$ group) which contains the ten parameter Galilean subgroup. This Galilean conformal group is to be contrasted with the twelve parameter Schr\"{o}dinger group (plus central extension) with which it has in common only the non-centrally extended Galilean subgroup. The Galilean conformal group is different from the Schr\"{o}dinger group in some crucial respects. For instance, the dilatation generator $\tilde{D}$  in the Schr\"{o}dinger group scales space and time differently 
$x_i \rightarrow \lambda x_i, t\rightarrow \lambda^2 t$. The corresponding generator $D$ in the Galilean Conformal Algebra (GCA) scales space and time in the {\it same} way $x_i \rightarrow \lambda x_i, t\rightarrow \lambda t$. Relatedly, the GCA does {\it not} admit a mass term as a central extension. Thus, in some sense, this symmetry describes "massless" non-relativistic theories, like the parent relativistic group but unlike the Schr\"{o}dinger group.

One of the most interesting feature of the GCA is its natural extension to an infinite dimensional symmetry algebra (which we will also often denote as GCA when there is no risk of confusion). This is somewhat analogous to the way in which the finite conformal algebra of $SL(2,C)$ in two dimensions extends to two copies of the Virasoro algebra. It is natural to expect this extended symmetry to be dynamically realized (perhaps partially) in actual systems possesing the finite dimensional Galilean conformal symmetry. This partial realization is actually observed in the non-relativistic Navier-Stokes equations. We refer you back to the paper \cite{Bagchi:2009my} for details.

It has been known (see \cite{Duval:1993pe} and references therein) that there is a notion of a "Galilean isometry"  which encompasses the so-called Coriolis group of arbitrary time dependent (but spatially homogeneous) rotations and translations.  In this language, our infinite dimensional algebra is that of "Galilean conformal isometries".  It contains one copy of a Virasoro together with an $SO(d)$ current algebra (on adding the appropriate central extension). Interestingly, an infinite extension for the Schr\"{o}dinger algebra has also been known for a long time \cite{Henkel:1993sg}, but has not been much exploited in the context of the non-relativistic limit of the AdS/CFT duality. See however \cite{Alishahiha:2009nm}

In this note, we would be looking at representations of this infinite dimensional GCA. The representations would be built along lines familiar in usual 2d CFTs. We would then focus on the explicit construction of the two and three point functions of the quasi-primary operators and comment on the differences of these correlation functions with those obtained in relativistic CFT and the non-relativistic Schr\"{o}dinger Algebra \cite{Henkel:1993sg}. 

The paper is organized as follows. In Sec2, we give a review of the essential features of the GCA which are worked out in more details in \cite{Bagchi:2009my}. We also review the infinite extension of the Schr\"{o}dinger Algebra following Henkel \cite{Henkel:1993sg}. In Sec3, we construct the representations of the GCA along the lines of usual relativistic CFT. In Sec4, we look at the correlation functions of the GCA. We explicitly work out the forms of the two and three point functions and then list the same for the Schr\"{o}dinger algebra as worked out in \cite{Henkel:1993sg} and point out the differences between the two and also the relativistic CFT. We end in Sec 5 with some comments and future directions.

\bigskip

{\bf{Note Added:}} Near the completion of this work, the paper \cite{Alishahiha:2009np} appeared on the arXiv. The authors, among other things, also address the question of the correlation functions of the GCA. We believe that their analysis is a special case of the more general results that we derive in this note. We comment more about this at the end of the paper.

\section{Non-relativistic Algebras}

\subsection{Galilean Conformal Algebras from Group Contractions}

The Galilean Conformal Algebra arises as a contraction of the parent relativistic conformal algebra. Physically this comes from taking the non-relativistic scaling 
\be{nrelscal}
t \rightarrow t \qquad   x_i \rightarrow \epsilon x_i
\ee
with $\epsilon \rightarrow 0$. This is equivalent to taking the velocities $v_i \sim \epsilon$ to zero
(in units where $c=1$).

Starting with the expressions for the Poincare generators ($\mu,\nu=0,1\ldots d$)
\be{galvec}
J_{\mu\nu} = -(x_{\mu} \p_{\nu} - x_{\nu} \p_{\mu}) \qquad P_{\mu}=\p_{\mu},
\ee
the above scaling gives us the Galilean vector field generators 
\begin{eqnarray}
J_{ij}&=& -(x_i\p_j-x_j\p_i) \qquad P_0=H= -\p_t \cr
P_i &=& \p_i \qquad J_{0i}=B_i= t\p_i .
\end{eqnarray}
This generates the Galilean sub-group of the GCA.
\ben{galalg}
[ J_{ij}, J_{rs} ] &=& so(d) \cr
[J_{ij} , B_r ] &=& -(B_i {\delta}_{jr} - B_j {\delta}_{ir}) \cr
[J_{ij},\, P_r] &=& -(P_i {\delta}_{jr} - P_j {\delta}_{ir}), \quad [J_{ij},\, H] = 0 \cr
 [B_i,B_j] &=& 0, \quad [P_i, P_j] =0 ,\quad [B_i, P_j] =0 \cr
[H, P_i] &=& 0, \quad [H, B_i] = - P_i. 
\een
The rest of the algebra follows from first scaling the relativistic dilatation and special conformal transformation generators.
\be{DKrel}
D = -(x\cdot\p) \qquad K_{\mu} = -(2x_{\mu}(x\cdot\p) -(x\cdot x)\p_{\mu})
\ee
where $D$ is the dilatation and $K_{\mu}$ are special conformal transformation generators.
The non-relativistic scaling in \eq{nrelscal} now gives
\ben{nrelconf}
D &=& -( x_i \p_i + t \p_t) \cr
K&=& K_0= -(2t x_i \p_i +t^2 \p_t) \cr
K_i &=& t^2\p_i. 
\een
The other non-trivial commutators of the GCA are \cite{Lukierski:2005xy}
\ben{galconalg}
[K, K_i] &=&0, \quad [K, B_i]=K_i, \quad [K, P_i]= 2B_i \cr
[J_{ij}, K_r] &=& -(K_i {\delta}_{jr} - K_j {\delta}_{ir}), \quad [J_{ij}, K] =0, \quad [J_{ij}, D]=0 \cr
[K_i,K_j] &=& 0, \quad [K_i, B_j]=0, \quad [K_i,P_j]=0, \quad [H, K_i] = -2B_i, \cr
[D, K_i] &=& -K_i, \quad [D , B_i]=0, \quad [D, P_i] = P_i,\cr
[D,H] &=& H,  \quad [H, K]= -2D, \quad [D, K]=-K.
\een

\subsection{The Infinite Dimensional Extended GCA}

The most interesting feature of the GCA is that it admits a very natural extension to an infinite 
dimensional algebra of the Virasoro-Kac-Moody type \cite{Bagchi:2009my}. To see this we denote 
\ben{rename}
L^{(-1)} &=& H, \qquad L^{(0)}=D, \qquad L^{(+1)}= K, \cr
M_i^{(-1)} &=& P_i, \qquad M_i^{(0)}=B_i, \qquad M_i^{(+1)}=K_i.
\een
The finite dimensional GCA which we had in the previous section can now be recast as
\ben{gcafinit}
[J_{ij}, L^{(n)}] &=&0 ,  \qquad [L^{(m)}, M_i^{(n)}] =(m-n)M_i^{(m+n)} \cr
[J_{ij} , M_k^{(m)} ] &=& -(M_i^{(m)} {\delta}_{jk} - M_j^{(m)} {\delta}_{ik}), 
\qquad [M_i^{(m)}, M_j^{(n)}] =0, \cr
[L^{(m)}, L^{(n)}] &=& (m-n)L^{(m+n)}.
\een
The indices $m,n=0,\pm 1$
We have made manifest the $SL(2,R)$ subalgebra with the generators $L^{(0)}, L^{(\pm 1)}$. 
In fact, we can define the vector fields 
\ben{gcavec}
L^{(n)} &=& -(n+1)t^nx_i\p_i -t^{n+1}\p_t \cr
M_i^{(n)} &=& t^{n+1}\p_i 
\een 
with $n=0,\pm 1$. These (together with $J_{ij}$) are then exactly the vector fields
in \eq{galvec} and \eq{nrelconf} which generate the GCA. 

If we now consider the vector fields of \eq{gcavec} for {\it arbitrary} integer $n$, and also define
\be{Jn}
J_a^{(n)} \equiv J_{ij}^{(n)}= -t^n(x_i\p_j-x_j\p_i)
\ee
then we find that this collection obeys the current algebra 
\ben{vkmalg}
[L^{(m)}, L^{(n)}] &=& (m-n)L^{(m+n)} \qquad [L^{(m)}, J_{a}^{(n)}] = -n J_{a}^{(m+n)} \cr
[J_a^{(n)}, J_b^{(m)}]&=& f_{abc}J_c^{(n+m)} \qquad  [L^{(m)}, M_i^{(n)}] =(m-n)M_i^{(m+n)}. 
\een
The index $a$ labels the generators of the spatial rotation group $SO(d)$ and $f_{abc}$ are the
corresponding structure constants. 
We see that the vector fields generate a $SO(d)$ Kac-Moody algebra without any central terms. In addition to the Virasoro and current generators we also have the commuting generators 
$M_i^{(n)}$ which function like generators of a global symmetry. The presence of these generators do not spoil the ability of the Virasoro-Kac-Moody generators to admit the usual central terms in their commutators.

\subsection{Schr\"{o}dinger Algebra}

The Schr\"{o}dinger symmetry group in $(d+1)$ dimensional spacetime has been more widely studied as a non-relativistic analogue of conformal symmetry. It's name arises from being the group of symmetries of the free Schr\"{o}dinger wave operator in $(d+1)$ 
dimensions. In other words, it is generated by those transformations that commute with the operator 
$S=i\p_t+{1\over 2m}\p_i^2$. However, this symmetry is also believed to be realized in 
interacting systems, most recently in cold atoms at criticality. 

The symmetry group contains the usual Galilean group and its algebra given by \refb{galalg} with the following central extension between momenta and boosts
\be{saextn}
[B_i, P_j] =m\delta_{ij} 
\ee 
The parameter $m$ is has the interpretation of being the non-relativistic mass (which also appears in the Schr\"{o}dinger operator $S$).

In addition to these Galilean generators there are {\it two} more generators which we will denote 
by $\tilde{K}, \tilde{D}$.  
$\tilde{D}$ is a dilatation operator, which unlike the relativistic case and the GCA, scales time and 
space differently. As a vector field $\tilde{D}= -(2t\p_t+x_i\p_i)$ so that
\be{tildact}
x_i \rightarrow \lambda x_i ,  \qquad t \rightarrow \lambda^2 t.
\ee
$\tilde{K}$ acts something like the time component of special conformal transformations. 
It has the form $\tilde{K} = -(tx_i\p_i+t^2\p_t)$ and generates the finite transformations 
(parametrised by $\mu$)
\be{tilkact}
x_i \rightarrow {x_i\over (1+\mu t)} ,\qquad  t \rightarrow {t \over (1+\mu t)}.
\ee

These two additional generators have non-zero commutators
\ben{schaddl}
[\tilde{K}, P_i] &=& B_i, \quad [\tilde{K}, B_i] = 0,  \quad [\tilde{D} , B_i]=-B_i  \cr 
[\tilde{D} , \tilde{K}] &=& -2\tilde{K}, \quad [\tilde{K}, H]= -\tilde{D}, \quad [\tilde{D}, H] = 2H.
\een
The generators $\tilde{K}, \tilde{D}$ are invariant under the spatial rotations $J_{ij}$.
We also see from the last line that $H, \tilde{K}, \tilde{D}$ together form an $SL(2,R)$ algebra. 
The central extension term of the Galilean algebra is compatible with all the 
extra commutation relations. 

A similar mass central extension between $P_i$ and $B_j$ cannot be added in the GCA because the Jacobi identities don't allow it.

Note that there is no analogue in the Schr\"{o}dinger algebra of the spatial components $K_i$ of special conformal transformations. 
Thus we have a  smaller group compared to the relativistic conformal group and the GCA. In $(3+1)$ dimensions
the Schr\"{o}dinger algebra has twelve generators (ten being those of the Galilean algebra) and the additional central  term. The relativistic conformal group and its non-relativistic contraction, the GCA have fifteen generators.

\subsection{Infinite extension of the Schr\"{o}dinger Algebra}

We wish to rewrite the Schr\"{o}dinger algebra in a different form and in a representation from which we would be able to see the infinite lift. 

The generators of the Schr\"odinger algebra can be thought of as vector fields defined on $d$ dimensional spacetime with the following representation
\ben{sarelabel}
&&J_{ij}=-(x_i\p_j-x_i\p_i),\quad P_i=-\p_i,\quad H=-\p_t,\quad m=-M\cr
&&B_i=-(t\partial_i+x_i M),\quad \tilde{D}=-(t\partial_t+\frac{1}{2}x_i\partial_i), \quad \tilde{K}=-(t^2\partial_t+tx_i\partial_i+\frac{1}{2} x^2 M).\nonumber
\een
Following \cite{Henkel:1993sg} one can define the generators of the corresponding infinite dimensional algebra in $d-1$ dimensions as follows
\ben{gen}
L_n&=&-t^{n+1}\partial_t-\frac{n+1}{2}t^{n} x_i\partial_i-\frac{n(n+1)}{4}t^{n-1} x^2 M ,\cr
Q_{i{\hat n}}&=&-t^{{\hat n}+1/2}\partial_i-({\hat n}+\frac{1}{2})t^{{\hat n}-1/2}x_iM,\quad T_n=-t^n M.
\een
Here $n\in Z$ and ${\hat n}\in Z+\frac{1}{2}$. It is straightforward to see
that the above generators satisfy the following commutation relations
\ben{SAinf}
&&[L_n,L_m]=(n-m)L_{n+m} \, ,\quad [Q_{i{\hat n}},Q_{j{\hat m}}]=({\hat n}-{\hat m})\delta_{ij}T_{{\hat n}+{\hat m}} \, ,\cr
&&[L_n,Q_{i{\hat m}}]=(\frac{n}{2}-{\hat m})Q_{i(n+{\hat m})} \, , \quad [L_n,T_m]=-mT_{n+m}.
\een
Due to non-trivial contribution of number operator to the Galilean boost the Schr\"odinger 
algebra does not allow an infinite dimensional extension for rotations $J_{ij}$, unlike the case of the GCA \cite{Alishahiha:2009nm}.

\section{Representations of the GCA}

Along the lines of usual relativistic conformal field theories, we can look to build the representations of the infinite dimensional algebra that we have obtained before. 

The representations are build by considering local operators which have well defined scaling dimension and ``boost" eigenvalue (which we will call the {\it{rapidity}}). We introduce the notion of local operators as 
\be{locop}
  \O(t,x) = U \O(0) U^{-1}, \quad \mbox{where} \quad U= e^{tH - x_i P^i } = e^{tL_{-1} - x_i M^i_{-1}}\,.
\ee
We note that $[L_0, M_0^i]=0$. The representations should thus be labeled by the eigenvalues of both the operators, as we stated before. Let us define local operators as those which are simultaneous eigenstates of $L_0, M^i_0$ :
\be{prim}
 [L_0, \O] = \D \O, \quad [M^i_0, \O] = \xi^i \O.
\ee

Let us look at the set of all local operators $\O_a(t,x)$.  These
operators, put at $t=0$ and $x=0$, form a representation of the contracted
subalgebra of the GCA which leaves the spacetime origin $\lbrace t=0,x=0 \rbrace$ invariant : for any operator $A$ in the subalgebra
\begin{equation}
  [A,\, \O_a(0)] = A_{ab} \O_b(0)\,.
\end{equation}

We would describe the irreducible representations of the infinite algebra.

We can use the Jacobi identities to show that 
\ben {l0ev}
[L_0, [L_n, \O]] &=& (\D-n) [L_n, \O] \,,\cr
[L_0, [M_n^i, \O]] &=& (\D-n) [M_n^i, \O]\,.
\een

The $L_n, M^i_n$ thus lower the value of the scaling dimension while $L_{-n}, M^i_{-n}$ raise it. Demanding that the dimension of the operators be bounded from below defines the primary operators in the theory to have the following properties :
\be{primop} 
[L_n, \O_p]= 0\,, \quad [M_n^i, \O_p] = 0\,,
\ee 
for all $n>0$. 

Starting with a primary operator $\O_p$, one can build up a tower of operators by taking commutators with $L_{-n}$ and $M^i_{-n}$. The operators built from a primary operator in this way form an irreducible representation of the contracted algebra. It is also possible to show that the full set of all local operators can be decomposed into irreducible representations, each of which is built upon a single primary operator. The task of finding the spectrum of dimensions of all local operators reduces to finding the $(\Delta \,,\,\xi^i )$ eigenvalues of the primary operators.

Curiously, the odd behaviour of the ``current algebra" commutation spoils the diagonalisation of this representation. To see this, let us look at how $M_0^i$ labels $[L_n, \O]$. Again, we use the Jacobi identity
\be{nondiag1} 
[M_0^i, [L_n, \O]] = \xi^i [L_n, \O] + n [M_n^i, \O]\,,
\ee 
which tells us that $[L_n, \O]$ is not an eigenstate of $M_0^i$. In a particular representation labelled by $(\D, \xi^i)$, when one moves up by acting with $L_n$, the significance of the rapidity, which is a vector quantum number, is not clear. 
But 
\be{diagM} 
 [M_0^i, [M^j_n, \O]] = \xi^i [M^j_n, \O] \,.
\ee 
So, moving up with $M^i_n$ preserves rapidity.\\
It should be possible to find a representation for which the action is diagonal. \\
Another point here is the unitarity of the representations. This is a problem which needs to be addressed if we are to use the full power of the infinite algebra to determine the form of the higher point correlation functions.

\section{Non-Relativistic Conformal Correlation Functions}

We wish to find the form of the correlation functions of the Galilean Conformal Algebras. To this end, let us define quasi-primary operators in this section.

We want to find the explicit form of the commutator $[L_n, \O(x,t)]$ for $\O(x,t)$ primary (i.e., obeying \refb{primop}) and $n \geq 0$.
\ben{ln}
&&[L_n, \O(x,t)] = [L_n , U \O(0) U^{-1}] \nonumber\\
&& \hspace{1cm}= [L_n,U]\O(0) U^{-1} + U \O(0) [L_n,U^{-1}] + U [L_n,\O(0) ]U^{-1} \\
&&\hspace{1cm} = U \lbrace U^{-1} L_n U - L_n \rbrace \O(0) U^{-1} + U \O(0) \lbrace L_n - U^{-1} L_n U \rbrace U^{-1} + \delta_{n,0}\Delta  \O(x,t)\,, \nonumber
\een
where $U$ is as defined in \refb{locop}.\\
Using the Baker-Campbell-Hausdorff (BCH) formula, we get
\ben{ljt} 
U^{-1} L_n \,U &=& \sum_{k=0}^{n+1}  {(n+1)! \over{(n+1-k)! k!}} (t^k L_{n-k} - kx_i t^{k-1} M^i_{n-k})\,, \cr
U^{-1} M^i_n \,U &=& \sum_{k=0}^{n+1}  {(n+1)! \over{(n+1-k)! k!}} t^k M^i_{n-k}\,. \nonumber
\een
Using the above, \refb{ln} gives us
\ben{ln1}
&& [L_n, \O(x,t)] = U [t^{n+1} L_{-1} - (n+1)t^n x_i M^i_{-1} , \O(0)] U^{-1}\cr
&& \hspace{2.7 cm} +\,U(n+1)(t^n \D - n t^{n-1} x_i \xi^i\O(0) U^{-1} \cr
&& = [t^{n+1} L_{-1} - (n+1)t^n x_i M^i_{-1} , \O(x,t)\,] + (n+1)(t^n \D - n t^{n-1} x_i \xi^i)\O(x,t). 
\een
Here, in the second line, we have used the fact that $U$ commutes with $L_{-1}$, $M^i_{-1}$ and any function of $(x,t)$.
Now $ [L_{-1}, \O(x,t)] = \p_t \O(x,t)\,,[M^i_{-1}, \O(x,t)] = -\p_i \O(x,t)\,$.
\\
Hence we finally obtain for $n \geq 0$ :
\be{lnfin}
[L_n, \O(x,t)] = [ t^{n+1}\p_t + (n+1)t^n x^i \p_i + (n+1)(t^n \D - n t^{n-1} x_i \xi^i)]\, \O(x,t)
\ee
By an analogous procedure, one can show that (again for $n \geq 0$)
\be{mn}
[M_n^i,\O(x,t)] =[ - t^{n+1}\p_i + (n+1)t^n \xi^i]\, \O(x,t).
\ee

\bigskip

Borrowing notation from the usual relativistic CFT, we would call those operators {\it{quasi-primary operators}} which transform as \refb{lnfin} and \refb{mn} with respect to the finite subalgebra $\{ L_0, L_{\pm 1}, M_0, M_{\pm 1}\}$.{\footnote{We do not look at the action of rotation because we would be labeling the operators with their boost eigenvalues, and boosts and rotations do not commute.}} We would examine the correlation functions of these quasi-primary operators. 

For any set of $n$ operators, one can define an $n$-point correlation
function,
\be{corrfun}
  G_n(t_1,x_1, t_2,x_2,....., t_n,x_n) = 
  \langle{0 | T \O_1(t_1,x_1) \O_2(t_2,x_2).....\O_n(t_n,x_n) |0}\rangle ,
\ee
where $T$ is time ordering.

The exponentiated version of the action of the dilatation operator is
\be{DOx-exp}
  e^{-\lambda L_0} \O(t,x) e^{\lambda L_0} = 
  e^{\lambda \D_\O} \O(e^{\lambda}t, e^\lambda x).
\ee
If all $\O_i$ have definite scaling dimensions then the correlation function \refb{corrfun} has a scaling property
\be{scl}
  G_n(e^{\lambda}t_i, e^\lambda x_i) = 
  \exp\left(-\lambda \sum_{i=1}^n \D_{\O_i}\right)
  G_n(t_i, x_i),
\ee
which follows from Eq.~(\ref{DOx-exp}) and $e^{\lambda L_0}|0\rangle=|0\rangle$.

\subsection{Two Point Function of the GCA}

We wish to first look at the details of how the two point function is constrained by the GCA. For that, we begin by considering two quasi-primary local operators $\O_1(x^i_1, t_1)$ and $\O_2(x^i_2,t_2)$ of conformal and rapidity weights $(\D_1, \xi^i_1)$ and $(\D_2, \xi^i_2)$ respectively.

The two point function is defined as
\be{2pt}
G^{(2)}(x_1^i, x_2^i, t_1, t_2) = \langle{0 | T \O_1(t_1,x^i_1) \O_2(t_2,x^i_2)|0}\rangle \,.
\ee 

We require the vacuum to be translationally invariant. The correlation functions can thus only depend on differences of the co-ordinates. From the fact that we are dealing with quasi-primaries, we get four more equations which would constrain the form of the 2-point function.

Let us label $\tau = t_1 - t_2$ and $r^i = x^i_1- x^i_2$ and call $\D= \D_1 + \D_2$ and $\xi^i = \xi^i_1 + \xi^i_2$. \\
We first look at the action under the Galilean boosts :
\ben{bst2pt}
&&\langle {0 | [M_0^i, G^{(2)} | 0} \rangle =0 \Rightarrow (-\tau \p_{r_i} + \xi_i)G^{(2)} =0 \cr
&\Rightarrow& G^{(2)} = C(\tau) \exp\left( {\xi_i r^i \over \tau} \right) \,,
\een
where $C(\tau)$ is an arbitrary function of $\tau$. \\
The behaviour under dilatations gives :
\be{dil2pt}
(\tau \p_{\tau} + r_i \p_{r_i} + \D ) G^{(2)}(r_i, \tau) = 0\,.
\ee 
This fixes the two point function to be 
\be{2}
G^{(2)}(r_i, \tau) = C^{(2)} \tau^{-\D} \exp\left( {\xi_i r^i \over \tau} \right)\,,
\ee 
where $C^{(2)}$ is an arbitrary constant. 

We still have two conditions left. These are :
\be{cond} 
\langle {0 | [L_1, G^{(2)}] | 0} \rangle =0, \quad \langle {0 | [M_1^i, G^{(2)}] | 0} \rangle =0\,,
\ee 
which give respectively
\be{hxi}
\D_1 = \D_2, \quad \xi^i_1 = \xi^i_2 \,.
\ee

So, in its full form, the two-point function for the GCA reads:
\be{2ptgca}
G^{(2)}(r_i, \tau) = C^{(2)} \delta_{\D_1,\D_2} \delta_{\xi_1, \xi_2} \tau^{-2\D_1} \exp\left( {2\xi^i_1 r_i \over \tau} \right)\,,
\ee 
where $C^{(2)}$ is, like before, an arbitrary constant.

\subsection{Three Point Function of the GCA}

Along lines similar to the previous subsection, we wish to construct the three point function of three quasi-primary operators $\O_a (\D=\D^a, \xi^i=\xi^a_i)$, where $a=1,2,3$.

The three point function is defined as
\be{3pt}
G^{(3)}(x_1^i, x_2^i, x_3^i, t_1, t_2, t_3) = \langle{0 | T \O_1(t_1,x^i_1)\O_2(t_2,x^i_2) \O_3(t_3,x^i_3)|0}\rangle \,.
\ee 

Again, remembering that we need to take only differences of co-ordinates into account, we define
$\tau = t_1 - t_3$, $\sigma = t_2 - t_3$ and $r^i = x^i_1- x^i_3$, $s^i = x^i_2- x^i_3$ and call $\D= \D^1 + \D^2 +\D^3$ and $\xi_i = \xi_i^1 + \xi_i^2 + \xi_i^3$.
The constraining equations are listed below.

Using the equation for dilatations, one obtains
\ben{dil3pt}
&&\langle{0|[L_0,\, T \O_{1}(\vec{x^{1}},t^{1}) \O_{2}(\vec{x^{2}},t^{2})) \O_{3}(\vec{x^{3}},t^{3})] |0}\rangle = 0\ \cr 
&\Rightarrow&  (\tau \partial_\tau +\sigma \partial_{\sigma} + r_{i}\p_{r_{i}}+s_{i}\p_{s_{i}}+\D )G^{(3)} = 0 .
\een
The boost equations give the following constraints :
\ben{bo3pt}
&& \langle{0|[M_0^{i},\, T \O_{1}(\vec{x^{1}},t^{1}) \O_{2}(\vec{x^{2}},t^{2})) \O_{3}(\vec{x^{3}},t^{3})] |0}\rangle=0\, \cr 
&\Rightarrow& (\tau \partial_{r_{i}}+\sigma \partial_{s_{i}} -\xi_{i} )G^{(3)} =  0\,.
\een
The non-relativistic spatial analogues of the special conformal transformations give
\ben{sct3pt}
&& \langle{0|[M_1^{i},\, T \O_{1}(\vec{x^{1}},t^{1}) \O_{2}(\vec{x^{2}},t^{2})) \O_{3}(\vec{x^{3}},t^{3})] |0}\rangle=0\, \cr 
&\Rightarrow& [(t^{a})^{2}\partial_{x^{a}_{i}} -2t\xi^{a}_i]G^{(3)} = 0 \,,
\een
which simplify to 
\be{simsct}
[\tau \sigma (\p_{r_i}+\p_{s_i})+\tau(\xi^1_i-\xi^{2}_i-\xi^{3}_i)+\sigma (-\xi^1_i+\xi^2_i-\xi^{3}_{i})]G^{(3)}=0 \,.
\ee
And finally, the last constraining equation comes from the temporal non-relativistic special conformal generator. 
Using $\langle{0|[L_1,\, T \O_{1}(\vec{x^{1}},t^{1}) \O_{2}(\vec{x^{2}},t^{2})) \O_{3}(\vec{x^{3}},t^{3})] |0}\rangle=0\,$, and equations for $L_0$ and $M_0^{i}$, one obtains :
\ben{tsct}
&[&\tau \sigma(\p_{\tau} + \p_{\sigma}) + (\sigma r_{i}+\tau s_{i})(\p_{r_{i}}+\p_{s_{i}})-\tau(\D^{1}-\D^{2}-\D^{3}) \cr 
&+& \sigma(\D^{1}-\D^{2}+\D^{3})+r_{i}(\xi^{1}_{i}-\xi^{2}_{i}-\xi^{3}_{i})-s_{i}(\xi^{1}_{i}-\xi^{2}_{i}+\xi^{3}_{i})]G^{(3)}=0.
\een

Motivated by the form of the the two-point function and examining \refb{bo3pt}, we make the following ansatz for the three-point function of the GCA :
\be{astz}
G^{(3)} =  f( \tau , \sigma , \D^a)  \exp\left( {\frac{a_{i}r^{i}}{\tau}}+ {\frac{b_{i}s^{i}}{\sigma}} +  {\frac{c_{i}(r^i - s^{i})}{(\tau - \sigma)}}    \right) \Psi(r_i, s_i, \tau, \sigma)\,, 
\ee
where $ f(\tau, \sigma , \D^a)$ is a function of $ \tau$, $\sigma$, and the $\D^a$'s; $\Psi(r_i, s_i, \tau, \sigma)$ is a function of $r_i, s_i, \tau, \sigma$, and independent of the $\D^a$'s and $\xi_i^a$'s; and $a_i, b_i, c_i$ (for $i = 1,2,3 $) are constants to be determined in terms of the $\xi_i^a$'s.

Plugging this form of $G^{(3)}$ in \refb{bo3pt} and remembering that $\Psi$ does not depend on the $\xi_i^a$'s, we get the constraints:
\be{xi}
a_i + b_i + c_i = \xi_i, \quad \mbox{and} \quad (\tau \partial_{r_{i}}+\sigma \partial_{s_{i}}) \Psi= 0.
\ee
Using \refb{dil3pt} and employing the $\D^a$-independence of $\Psi$, we get the equations :
\be{di}
(\tau \partial_\tau +\sigma \partial_{\sigma} ) f(\tau, \sigma, \D^a) = -\D\,,
\ee{}
and
\be{cons2}
(\tau \partial_\tau +\sigma \partial_{\sigma} + r_{i}\p_{r_{i}}+s_{i}\p_{s_{i}}) \Psi = 0\,.
\ee
Eq.\refb{di} dictates that $f(\tau, \sigma , \D^a)$ should have the form
\be{f}
f(\tau, \sigma,\D^a) = \tau ^{-A} \sigma^{-B} (\tau-\sigma)^{-C}\,,
\ee{}
where $A, B, C$ depend on the $\D^a$'s and obey the condition:
\be{h}
A+B+C = \D\,.
\ee{}
Now the constraints in Eq.\refb{simsct} fix $a_i, b_i$ to be
\be{ab}
a_i = \xi_i^1 + \xi_i^3 -\xi_i^2 \hspace{2mm}, \quad b_i = \xi_i^2 + \xi_i^3 -\xi_i^1\,,
\ee
and restrict $\Psi$ to obey :
\be{cons3}
\tau \sigma (\p_{r_i}+\p_{s_i})\Psi= 0\,.
\ee
Finally, Eq.\refb{tsct} fixes $A, B$ to be
\be{ABh}
A = \D^1+\D^3-\D^2 \hspace{2mm}, \quad  B= \D^2+\D^3-\D^1\,,
\ee{}
and constrains $\Psi$ to obey :
\be{cons4}
\lbrace \tau \sigma(\p_{\tau} + \p_{\sigma}) + (\sigma r_{i}+\tau s_{i})(\p_{r_{i}}+\p_{s_{i}})) \rbrace  \Psi=0 \,.
\ee
Using \refb{xi}, \refb{ab}, \refb{h} and \refb{ABh}, the remaining constants $c_i, C$ are found to be
\be{cC}
c_i = \xi_i^1 + \xi_i^2 -\xi_i^3 \hspace{2mm}, \quad  C= \D^1+\D^2-\D^3\,.
\ee{}
The equations \refb{xi} and \refb{cons3} make $\Psi$ a function of $\tau$ and $\sigma$ alone. Then applying \refb{cons2} and \refb{cons4}, in a similar way, we find that $\Psi$ is restricted to a constant.

Finally, the full three point function is:
\ben{3ptgca}
&& G^{(3)}(r_i, s_i, \tau, \sigma) \cr 
&& = C^{(3)} \tau^{-(\D^{1}-\D^{2}+\D^{3})} \sigma^{-(\D^{2}+\D^{3}-\D^{1})} (\tau-\sigma)^{-(\D^{1}+\D^{2}-\D^{3})} \cr 
&& \hspace{5 mm}\times \exp\left( {\frac{(\xi^{1}_{i}+ \xi_i^3 -\xi_i^2)r^{i}}{\tau}}+ {\frac{(\xi^{2}_{i}+ \xi_i^3 -\xi_i^1) s^{i}}{\sigma}}  + {\frac{(\xi^{1}_{i}+ \xi_i^2 -\xi_i^3) (r^i-s^{i})}{(\tau-\sigma)}}\right) \,,\nonumber \\ 
\een
where $C^{(3)}$ is an arbitrary constant.

So we see that like in the case of relativistic CFTs, the three point function is fixed upto a constant.

\subsection{Comparing Correlation Functions}

We want to understand the expressions for the two and three point functions better. To that end, we wish to recapitulate the two and three point functions in the usual relativistic CFTs and the Schr\"{o}dinger algebra.

As is very well known, in the relativistic CFTs in any dimension, the conformal symmetry is enough to fix the two and three point functions upto constants. Framed in the language of 2D CFTs, one needs only the finite $SL(2, R)$ sub-group of the full Virasoro algebra to fix the 2 and 3 point function. Let us list them: \\
Two point function:
\be{cft2pt} 
G^{(2)}_{CFT}(z_i, \overline{z_i}) = \delta_{h_1, h_2} \delta_{\overline{h}_1, \overline{h}_2} {C_{12} \over {z_{12}^{2h} \overline{z}_{12}^{2 \overline{h}}}}
\ee
Three point function:
\ben{cft3pt} 
G^{(3)}_{CFT}(z_i, \overline{z_i}) &=& C_{(123)} z_{12}^{-(h_1+h_2-h_3)} z_{23}^{-(h_2+h_3-h_1)} z_{13}^{-(h_3+h_1-h_2)} \nonumber \\
 & & \times \mbox{non-holomorphic}
\een 

One can find the two-point and three point functions for the quasi-primary operators in the Schr\"{o}dinger Algebra as well \cite{Henkel:1993sg}. We list the answers here: \\
Two point function:
\be{sa2pt} 
G^{(2)}_{SA} = C_{(12)} \delta_{\D_1, \D_2}\delta_{m_1, m_2} (t_1-t_2)^{\D_1} \exp \left( { m_1{(x_1^i - x_2^i)}^2 \over {2 (t_1- t_2)}} \right) 
\ee 
Three point function:
\ben{sa3pt} 
G^{(3)}_{SA} &=& C_{(123)} \delta_{m_1+m_2, m_3} (t_1-t_3)^{-(\D_1+\D_3-\D_2)\over{2}} (t_2-t_3)^{-(\D_2+\D_3-\D_1)\over{2}} \cr 
 & & (t_1-t_2)^{-(\D_1+\D_2-\D_3)\over{2}} \exp \left( { m_1(x_1^i - x_3^i)^2 \over {2(t_1- t_3)}} + { m_2 (x_2^i - x_3^i)^2 \over{(t_2-t_3)}}\right) \cr 
&\times& \Psi \left( {{[(x^i_1-x^i_3)(t_2-t_3) - (x^i_2-x^i_3)(t_1-t_3)]}^2}\over{(t_1-t_2)(t_2-t_3)(t_1-t_3)}\right)  
\een

\bigskip

Some remarks are in order. We first note that all the two-point functions vanish if the quasi-primaries are of different weight. The non-relativistic ones are further constrained by the equality of the rapidity and mass eigenvalues for the GCA and SA respectively.\\
The exponential part is an inherently non-relativistic effect. This form comes from the generator of Galilean boosts. \\
The GCA and SA three point functions {\it{crucially differ}}. As we have seen above, the GCA one is completely fixed upto a constant factor, like in usual CFT. But for the Schr\"{o}dinger symmetry, the three point function is fixed only upto an arbitrary function of a particular combination of variables. This is a result of the differing number of generators. The finite sub-algebra of the GCA has two more generators and this constrains the form of the three-point function further. \\
The Schr\"{o}dinger algebra correlators also have an additional superselection rule of the mass. But there is no analogue of such a selection rule for the GCA. This is because the boosts, unlike the mass, acts non-trivially on the co-ordinates. The action of the boosts in this case is more like the dilation operator which also has a non-trivial action on the space-time.{\footnote{We would like to thank Shiraz Minwalla for pointing this out and Ashoke Sen for valuable discussions on this point.}}

\section{Concluding Remarks}

In \cite{Bagchi:2009my}, a bulk dual to the Galilean Conformal Algebra was proposed. This has the form of a vector bundle, where we have $AdS_2$ as a base and there are $R^d$ fibres on this base. The gravitational theory on this space can be formulated in a spirit similar to the Newton-Cartan description of the non-relativistic limit of Einstein gravity \footnote{See e.g. \cite{Misner:1974qy} for a discussion of the standard Newton-Cartan theory.}. 
The metric in this language degenerates, and we can no longer speak of a single metric for the whole of the spacetime. There exists a metric $g_{\alpha \beta}$ on the $AdS_2$ and a flat metric $\delta_{ij}$ on the spatial slices. The dynamic degrees of freedom are the non-metric connections $\Gamma^{\rho}_{\mu\nu}$ which talk between the two metrics. The GCA emerges as the asymptotic isometries \cite{Brown:1986nw} of this Newton-Cartan like description.
Taking the non-relativistic scaling limit of metric on the Poincare patch in $AdS_{d+2}$ leaves one with the $AdS_2$ metric.
In this light, we had talked about an $AdS_2/CFT_1$ description in the paper which seems to exist as the non-relativistic limit of all the $AdS/CFT$ dualities. 
Similar $AdS_2$ factors have also been noted while realising the Schr\"{o}dinger Algebra as asymptotic isometries in \cite{Alishahiha:2009nm}. So, it is plausible that all non-relativistic limits would isolate some sort of an $AdS_2/CFT_1$ duality. 

In the correlation functions obtained above and those which have been obtained from the Schr\"{o}dinger algebra there is further evidence of this $AdS_2/CFT_1$. Let us explain how. 

We have seen that both the two and three point function in the cases of the GCA and the Schr\"{o}dinger algebra contains pieces which are reminiscent of the parent relativistic theory, but those pieces have been the pieces where only the time co-ordinate was present. The $AdS_2/CFT_1$ that we talk about separates out the radial $z$ and the temporal $t$ direction in the bulk. This time co-ordinate would be the one which leads to the $CFT_1$. What we get from the field theory side is precisely this conformal behaviour of the time piece. The exponential which contains both the time and space dependent terms probably arises out of the complicated fibre-bundle structure. The two non-relativistic theories would be realised as different fibre structures in the bulk and possibly, this is why there are differences in the exponential pieces. 

There have been interesting new developments in \cite{Fuertes:2009ex, Volovich:2009yh}, where the authors find the correlation functions of the Schr\"{o}dinger algebra from the bulk using a bulk-boundary dictionary. It would be useful to do a similar analysis in the case of the GCA but since we don't yet have a bulk description in terms of a non-degenerate metric, the bulk-boundary dictionary might prove more tricky. One possible way out might be to look at a realization in two higher dimension instead of one, like the bulk dual of the Schr\"{o}dinger algebra and use that as a starting point. These points are under investigation. 

Near the end of this project, the paper \cite{Alishahiha:2009np} appeared on the arXiv. We stated in the introduction that their correlation functions are a specialized case of our results. The difference arises from the fact that the states of \cite{Alishahiha:2009np} have been implicitly (?) chosen to have zero rapidity eigenvalue. As we have argued before, the fact that $[D, B_i]=0$ or in redefined language $[L_0, M^i_0]=0$ means that one necessarily needs to look at states/operators which are simultaneous eigenvectors of these two generators. If we choose to look at a sector where the rapidity eigenvalue for all fields is zero, then from Eq.\refb{2ptgca} and Eq.\refb{3ptgca} it is easy to see that our correlation functions reproduce the expressions obtained in \cite{Alishahiha:2009np}. The fact that there is more structure in the correlation functions is a reflection of the fact that the naive scaling of the correlation functions of the relativistic CFT would not work if one wishes to reproduce the expressions \refb{2ptgca}, \refb{3ptgca}. The zero rapidity sub-sector seems to simplify the analysis and it might be interesting to understand what is the physical significance of looking at this limit of the GCA representations.

\subsection*{Acknowledgements}
The authors would like to thank Rajesh Gopakumar for many helpful discussions and encouragement and for carefully reading and commenting on the manuscript. We would also like to thank Shamik Banerjee, Bobby Ezhuthachan, Shiraz Minwalla, Ashoke Sen for discussions and Sangmin Lee for initial collaboration. One of us (A.B.) would like to thank the string theory group at the ETH, Z\"{u}rich where the work was presented. We would like to thank ICTP, Trieste for hospitality during the completion of this work. We would also like to record our indebtedness to the people of India for their generous support for fundamental enquiries.



\begin{thebibliography}{999}


\bibitem{Maldacena:1997re}
  J.~M.~Maldacena,
  ``The large N limit of superconformal field theories and supergravity,''
  Adv.\ Theor.\ Math.\ Phys.\  {\bf 2}, 231 (1998)
  [Int.\ J.\ Theor.\ Phys.\  {\bf 38}, 1113 (1999)]
  [arXiv:hep-th/9711200].

\bibitem{Bagchi:2009my}
  A.~Bagchi and R.~Gopakumar,
  ``Galilean Conformal Algebras and AdS/CFT,''
  arXiv:0902.1385 [hep-th].

\bibitem{Hartnoll:2009sz}
  S.~A.~Hartnoll,
  ``Lectures on holographic methods for condensed matter physics,''
  arXiv:0903.3246 [hep-th].

\bibitem{Nishida:2007pj}
  Y.~Nishida and D.~T.~Son,
  ``Nonrelativistic conformal field theories,''
  Phys.\ Rev.\  D {\bf 76}, 086004 (2007)
  [arXiv:0706.3746 [hep-th]].

\bibitem{Hagen:1972pd}
  C.~R.~Hagen,
  ``Scale and conformal transformations in galilean-covariant field theory,''
  Phys.\ Rev.\  D {\bf 5}, 377 (1972).

\bibitem{Niederer:1972zz}
  U.~Niederer,
  ``The maximal kinematical invariance group of the free Schrodinger
  equation,''
  Helv.\ Phys.\ Acta {\bf 45}, 802 (1972).


\bibitem{Henkel:1993sg}
  M.~Henkel,
  ``Schr\"odinger invariance in strongly anisotropic critical systems,''
  J.\ Statist.\ Phys.\  {\bf 75}, 1023 (1994)
  [arXiv:hep-th/9310081].

\bibitem{Son:2005rv}
  D.~T.~Son and M.~Wingate,
  ``General coordinate invariance and conformal invariance in nonrelativistic
  physics: Unitary Fermi gas,''
  Annals Phys.\  {\bf 321}, 197 (2006)
  [arXiv:cond-mat/0509786].

\bibitem{Son:2008ye}
  D.~T.~Son,
  ``Toward an AdS/cold atoms correspondence: a geometric realization of the
  Schroedinger symmetry,''
  Phys.\ Rev.\  D {\bf 78}, 046003 (2008)
  [arXiv:0804.3972 [hep-th]].

\bibitem{Balasubramanian:2008dm}
  K.~Balasubramanian and J.~McGreevy,
  ``Gravity duals for non-relativistic CFTs,''
  Phys.\ Rev.\ Lett.\  {\bf 101}, 061601 (2008)
  [arXiv:0804.4053 [hep-th]].

\bibitem{Goldberger:2008vg}
  W.~D.~Goldberger,
  ``AdS/CFT duality for non-relativistic field theory,''
  arXiv:0806.2867 [hep-th].

\bibitem{Barbon:2008bg}
  J.~L.~B.~Barbon and C.~A.~Fuertes,
  ``On the spectrum of nonrelativistic AdS/CFT,''
  JHEP {\bf 0809}, 030 (2008)
  [arXiv:0806.3244 [hep-th]].

\bibitem{Herzog:2008wg}
  C.~P.~Herzog, M.~Rangamani and S.~F.~Ross,
  ``Heating up Galilean holography,''
  arXiv:0807.1099 [hep-th].

\bibitem{Maldacena:2008wh}
  J.~Maldacena, D.~Martelli and Y.~Tachikawa,
  ``Comments on string theory backgrounds with non-relativistic conformal
  symmetry,''
  JHEP {\bf 0810}, 072 (2008)
  [arXiv:0807.1100 [hep-th]].

\bibitem{Adams:2008wt}
  A.~Adams, K.~Balasubramanian and J.~McGreevy,
  ``Hot Spacetimes for Cold Atoms,''
  arXiv:0807.1111 [hep-th].


\bibitem{Alishahiha:2009nm}
  M.~Alishahiha, R.~Fareghbal, A.~E.~Mosaffa and S.~Rouhani,
  ``Asymptotic symmetry of geometries with Schrodinger isometry,''
  arXiv:0902.3916 [hep-th].

\bibitem{Fuertes:2009ex}
  C.~A.~Fuertes and S.~Moroz,
  ``Correlation functions in the non-relativistic AdS/CFT correspondence,''
  arXiv:0903.1844 [hep-th].


\bibitem{Volovich:2009yh}
  A.~Volovich and C.~Wen,
  ``Correlation Functions in Non-Relativistic Holography,''
  arXiv:0903.2455 [hep-th].

\bibitem{Lukierski:2005xy}
  J.~Lukierski, P.~C.~Stichel and W.~J.~Zakrzewski,
  ``Exotic Galilean conformal symmetry and its dynamical realisations,''
  Phys.\ Lett.\  A {\bf 357}, 1 (2006)
  [arXiv:hep-th/0511259].



\bibitem{BMN}
  D.~E.~Berenstein, J.~M.~Maldacena and H.~S.~Nastase,
  ``Strings in flat space and pp waves from N = 4 super Yang Mills,''
  JHEP {\bf 0204}, 013 (2002)
  [arXiv:hep-th/0202021].

\bibitem{Duval:1993pe}
  C.~Duval,
  ``On Galilean isometries,''
  Class.\ Quant.\ Grav.\  {\bf 10}, 2217 (1993).


\bibitem{Brown:1986nw}
  J.~D.~Brown and M.~Henneaux,
  ``Central Charges in the Canonical Realization of Asymptotic Symmetries: An
  Example from Three-Dimensional Gravity,''
  Commun.\ Math.\ Phys.\  {\bf 104}, 207 (1986).

\bibitem{Alishahiha:2009np}
  M.~Alishahiha, A.~Davody and A.~Vahedi,
  ``On AdS/CFT of Galilean Conformal Field Theories,''
  arXiv:0903.3953 [hep-th].

\bibitem{Misner:1974qy}
  C.~W.~Misner, K.~S.~Thorne and J.~A.~Wheeler,
 ``Gravitation,''
{\it  San Francisco 1973, 1279p}

\end{thebibliography}
\end{document}